# Experimental evidence for orbital magnetic moments generated by moiré-scale current loops in twisted bilayer graphene

Si-Yu Li[1], Yu Zhang[1], Ya-Ning Ren[1], Jianpeng Liu[2,†], Xi Dai[2], and Lin He[1,3,†]

[1] Center for Advanced Quantum Studies, Department of Physics, Beijing Normal University, Beijing, 100875, People's Republic of China

[2] Department of Physics, Hong Kong University of Science and Technology, Kowloon, Hong Kong, China

[3] State Key Laboratory of Functional Materials for Informatics, Shanghai Institute of Microsystem and Information Technology, Chinese Academy of Sciences, 865 Changning Road, Shanghai 200050, People's Republic of China

†Correspondence and requests for materials should be addressed to J.L. (e-mail: jianpeng@ust.hk) and L.H. (e-mail: helin@bnu.edu.cn).

**A remarkable property of twisted bilayer graphene (TBG) with small twist angle is the presence of a well-defined and conserved low-energy valley degrees of freedom[1], which can potentially bring about new types of valley-associated spontaneous-symmetry breaking phases. Electron-electron ($e$-$e$) interactions in the TBG near the magic angle ~ 1.1 ° can lift the valley degeneracy, allowing for the realization of orbital magnetism and topological phases[2-11]. However, direct measurement of the orbital-based magnetism in the TBG is still lacking up to now. Here we report evidence for orbital magnetic moment generated by the moiré-scale current loops in a TBG with a twist angle $\theta$ ~ 1.68 °. The valley degeneracy of the 1.68 ° TBG is removed by $e$-$e$ interactions when its low-energy van Hove singularity (VHS) is nearly half filled. A large and linear response of the valley splitting to magnetic fields is observed, attributing to coupling to the large orbital magnetic moment induced by chiral current loops circulating in the moiré pattern. According to our experiment, the orbital magnetic moment is about 10.7 $\mu_B$ per moiré supercell. Our result paves the way to explore magnetism that is purely orbital in slightly twisted graphene system.**

Owning to strong electron-electron (*e-e*) interactions, twisted bilayer graphene (TBG) near the magic angle ~ 1.1 ° exhibits many interesting correlation phases, including insulating, superconducting, ferromagnetic, and topological phases[9-14]. One of the most remarkable properties in the magic-angle TBG is that the low-energy bands can have non-zero valley Chern numbers[2-11], which not only allow for the possibility of topological phases, but also enable the realization of the first example of a material with purely orbital-based magnetism. Although (quantum) anomalous Hall effect, *i.e.*, the topological phase, has been observed very recently in the magic-angle TBG[9-11], direct measurement of the orbital-based magnetism is still lacking. Theoretically, the orbital magnetic moment in the magic-angle TBG is generated by chiral current loops on the moiré scale (~ 10 nm), thus it is expected to be much larger than that on the atomic lattice scale[2-11,15]. In this work, we report evidence for giant orbital magnetic moment generated by moiré-scale current loops in a 1.68 ° TBG with the period of the moiré pattern ~ 8.4 nm. Our high-resolution scanning tunneling microscope and spectroscopy (STM and STS) measurements demonstrate that the *e-e* interactions lift the valley degeneracy of the TBG when the low-energy flat bands are half filled such that the chemical potential is around the van Hove singularity (VHS). The valley splitting of the VHS increases linearly with increasing magnetic fields due to the coupling to the orbital magnetic moment, which is deduced to be about 10.7 $\mu_B$ per moiré supercell in our experiment. Our analysis indicates that the $C_{2z}$ symmetry breaking of the TBG induced by the substrate is crucial for nonvanishing valley Chern number, resulting in the emergence of the large orbital magnetic moment in the studied system.

In our experiment, twisted multilayer graphene (TMG) was synthesized on Rhodium (Rh) foils via a traditional ambient pressure chemical vapor deposition method. According to previous studies[16-19], the graphene layers grown on the Rh foils have a strong twisting tendency, providing rich TMG with different twist angles. Figure 1a shows a representative STM topographic image of a TMG region, which contains two moiré patterns with different periods (see inset for fast Fourier transforms of the STM image). The small period of the moiré pattern is $D ≈ 1.5$ nm and the corresponding twist angle between two adjacent graphene sheets is estimated as about $\theta ≈ 9.1±0.2°$ according to $D = a/[2\sin(\theta/2)]$, where $a \approx 0.246$ nm is the lattice constant of graphene. The periods of the large moiré pattern exhibit obvious anisotropic and they are measured as $L_1 = 8.50$ nm, $L_2 = 8.63$ nm, and $L_3 = 8.05$ nm, as shown in Fig. 1a. The slightly deformed moiré pattern indicates that there is heterostrain in the topmost TBG. With assuming an uniaxial strain in the TBG[20-25], the twist angle of the large-period moiré pattern can be estimated as $\theta \approx 1.68°$ with the uniaxial heterostrain $\varepsilon \approx -0.2\%$ along 23.5 °with respect to the *x* axis. To further determine the stacking order of the topmost three layers, we measured atomic-resolution STM images of the AB-stacked region (Fig. 1b), AA-stacked region (Fig. 1c), and BA-stacked region (Fig. 1d) in the large-period moiré pattern. We can clearly observe hexagonal honeycomb lattices in the AA-stacked region and triangular lattices in the AB/BA stacked region, which are exactly consistent with the features observed in slightly TBG[20-29], indicating that the large-period moiré pattern arises from the twist between the first and the second graphene sheets. Figure 1e shows a schematic image of the topmost three graphene layers in Fig. 1a. The large twist angle between the second and the third layer $\theta_{23} \approx 9.1°$, which enables that the topmost TBG is electronically decoupled from the supporting substrate[18,19,34,35]. Therefore, the probed electronic states by the STM should be mainly contributed by the topmost TBG with $\theta_{12} \approx 1.68°$, which is demonstrated explicitly in our experiment, as illustrated subsequently.

Figure 1f shows representative d$I$/d$V$ spectra recorded in the AA-stacked region and the AB/BA stacked region of the topmost TBG. We can make three obvious observations from the spectra. The first is that the spectrum recorded in the AA regions of the moiré pattern shows low-energy sharp peaks, which are the VHSs of the topmost TBG. However, the VHSs exhibit four peaks rather than two peaks as expected in pristine TBG[20-29]. Such a feature arises from the heterostrain of the topmost TBG and will be elaborated in Fig. 2. The second notable feature of the spectra is the presence of the step-like features, which are mainly localized on AB/BA-stacked region, at higher and lower energies, as marked by arrows in Fig. 1f. Similar step-like features are also observed in the tunneling spectra of the magic-angle TBG and are also generated by the hetrostrain[21]. The third feature in the spectrum recorded in the AA regions is the appearance of negative differential conductance (NDC) between the low-energy VHSs and high-energy step-like features, as pointed out in Fig. 1f. Previous studies demonstrated that the existence of an energy gap is necessary to observe the NDC in the tunneling spectra of graphene systems[30,31]. Therefore, the observed NDC is a clear signature that there is a gap between the low-energy VHSs and the remote bands in the topmost TBG, which agrees well with that observed in slightly TBG[9-14,20-23,32,33]. These features demonstrated explicitly that the tunneling spectra in our studied system are mainly contributed by the topmost TBG with $\theta_{12} \approx 1.68°$.

To further explore the electronic properties, we carried out high-resolution STS measurements in the topmost TBG with different doping (the energy resolution is about 1 meV, see Supplemental materials for details of the method to change the doping of the topmost TBG). Figure 2a shows a representative high-resolution STS spectrum recorded in the AA regions of the topmost TBG where the VHSs are completely empty. The charge neutrality point $n_c$ of the topmost TBG is about 50 meV above the Fermi energy and, obviously, there are four low-energy VHSs. Usually, there are only two VHSs flanking the $n_c$ of the slightly TBG[20-29]. Our result indicates that each VHS of the topmost TBG splits into two peaks. When the VHSs are partially filled, as shown in Fig. 2b, the *e-e* interactions further split the left two VHSs into four peaks, leaving the right two unoccupied VHSs almost unchanged. Therefore, the *e-e* interactions should not be the reason for the emergence of four VHSs when they are completely empty (Fig. 2a). We attribute this phenomenon to the existence of heterostrain in the topmost TBG, which is clearly shown by the three different moiré lattice constants in Fig. 1a. The direction and magnitude of the strain in the topmost TBG can be uniquely determined by the three non-equivalent moiré lattice constants. With assuming only the top layer graphene is strained and the other graphene layers are undistorted, we can obtain the low-energy band structures and local density of states (DOS) of the twisted graphene systems, as summarized in Fig. 2c-2f (see Supplemental materials for details). Without strain, there are two VHSs flanking the $n_c$ of the 1.68 °TBG, as shown in Fig. 2f. By introducing the heterostrain, each VHS splits into two, which is consistent with the experimental data. Even with considering the Bernal-stacked bilayer graphene as a supporting substrate of the 1.68 °TBG, each VHS will still be split into two peaks by the heterostrain, as shown in Fig. 2d. However, the existence of the bilayer graphene as a substrate will remove the gap between the low-energy VHSs and the remote bands, as shown in Fig. 2c and 2d. Such a feature is not consistent with our experimental result, further confirming that the tunneling spectra in our studied system are mainly contributed by the topmost TBG. However, we should point out that the presence of the multilayer graphene as a substrate is still crucial since it breaks the $C_{2z}$ symmetry of the topmost TBG, which allows for non-vanishing Berry curvatures and the associated phenomena.

Now we begin to discuss the *e-e* interactions in the topmost TBG. When the doping of the TBG varies about 20 meV and the left two VHSs are half filled, the *e-e* interactions split the left two VHSs into four peaks, as shown in Fig. 2b (the splitting is about 42 meV). Simultaneously, the energy separation between the VHSs below the $n_c$ and above the $n_c$ is increased as compared to the case that the four VHSs are completely empty (the energy separation increases from about 50 meV to about 60 meV when the VHSs are partially filled). These two features are also observed in the magic-angle TBG and indicate that the *e-e* interactions are also very important in the 1.68 °TBG even though the twist angle is larger than the magic angle by about 52%. Therefore, it is possible to realize interesting correlated phenomena in the TBG with a wide range of twist angle[14,36,37]. Such a result also indicates that the graphene multilayer substrate poorly screens the *e-e* interactions in the TBG. An examination of the spectrum shows that the intensities of the split VHSs on either side of the Fermi level are about half the intensity of the un-split VHSs (Fig. 2b), indicating that either the spin or the valley degeneracy of the TBG is lifted and the obtained state is twofold degeneracy. Previous transport measurements also observe twofold degeneracy of the Landau fan in the magic-angle TBG when the VHS is half filled[14].

To further explore the nature of the splitting at half filling of the VHSs in the 1.68 ° TBG, we carried out high-resolution STS measurements in the presence of different perpendicular magnetic fields, as summarized in Fig. 3. When the VHSs are completely empty, the spectra are almost un-affected by the magnetic fields. However, when the

VHSs are half filled, the splitting increases quickly with increasing the magnetic fields, as shown in Fig. 3b. The half-filled splitting $\Delta E$ as a function of magnetic field is plotted in Fig. 3c, which exhibits a linear scaling. Derived from the slope of the linear fits, we obtain an effective $g$ factor as about 21.4, which is much larger than that of the Zeeman splitting of spin (the effective $g$ factor of spin is usually about 2). In fact, the obtained giant $g$ factor is quite similar as that of the valley splitting when the zero Landau level of graphene monolayer is half filled[18,38]. In slightly TBG, when the valley degeneracy is lifted by the $e$-$e$ interactions and the $C_{2z}$ symmetry is broken by the substrate, each valley would be associated with non-vanishing Berry curvature[2-11]. Then, there is large orbital magnetic moment generated by circulating current loops in the moiré patterns of the TBG (Figure 4a). According to the obtained slope in Fig. 3c, the orbital magnetic moment in each moiré is estimated as about 10.7 $\mu_B$. Below, we will show that the existence of large orbital magnetic moment in the TBG is the origin of the large response of the splitting to the magnetic fields.

Before carrying out quantitative calculation, we first review the symmetries of the TBG and the necessary conditions for the emergence of orbital magnetic moment in the TBG. The TBG system possesses $C_{3z}$, $C_{2y}$, $C_{2x}$, $C_{2z}$, and $T$ (time-reversal) symmetries. As both $C_{2z}$ and $T$ are symmetry operations that interchange the two valleys, the combined symmetry operation $C_{2z}T$ is a symmetry associated with each valley, which enforces the Berry curvature to be vanishing at every $k$ point in the moiré Brillouin zone. As a result, all the Berry-curvature-related phenomena such as orbital magnetism and anomalous Hall effect would be forbidden by such $C_{2z}T$ symmetry even in the presence of finite valley polarizations. If the $C_{2z}$ symmetry of the TBG is broken by the substrate, for example, the TBG is aligned with hexagonal boron nitride (hBN) substrate, then each valley would be associated with non-vanishing Berry curvature due to the $C_{2z}T$ symmetry breaking. Such $C_{2z}T$ symmetry breaking is crucial in achieving phenomena such as quantum anomalous Hall effect and orbital ferromagnetism. For example, in continuously twisted trilayer system, $C_{2z}T$ is still an emergent symmetry of the continuum Hamiltonians, which kills Berry curvature and the associated phenomena. In our experiment, the simplest way to break the $C_{2z}$ symmetry is assume that the Bernal-stacked bilayer graphene is below the topmost TBG as a substrate. Then, there is no $C_{2z}T$ symmetry for each valley, which allows for nonzero Berry curvatures without the necessity of an hBN substrate. In Fig. 2c, the calculated band structures of the TBG supported by the Bernal-stacked bilayer graphene possess nonzero valley Chern numbers, which resembles the case of TBG with the staggered sublattice potential, *i.e.*, the band structures of hBN-aligned TBG (Fig. 2e). The DOS spectra of the two cases are also quite similar, as shown in Fig. 2d and 2f. Therefore, we believe that the Coulomb interactions in our system would have similar effects in hBN-aligned TBG. The difference is that the $C_{2z}T$ symmetry is broken by the multilayer graphene substrate in our experiment, rather than by the hBN in the hBN-aligned TBG.

Based on the above argument, we have carried out a self-consistent unrestricted Hartree-Fock calculation for the TBG with the staggered sublattice potential. At -1/2 filling of the two low-energy flat bands, the chemical potential is around the VHS on the hole side. As the Hartree-Fock calculation is fully unrestricted, all the valley, spin, and sublattice dependence of the mean-field order parameters has been taken into account, and the details of such moiré Hartree-Fock method are presented in Ref. 8. The calculated mean-field ground state is a valley-polarized state, as shown in Fig. 4b, which is expected to possess chiral current loops. With assuming that the K valley is 100% polarized due to spontaneous valley symmetry breaking from coulomb interactions, as observed in our experiment, we can obtain the real-space distributions of the local current density and the current-induced local magnetic field within the moiré Wigner-Seitz cell for the TBG, as shown in Fig. 4a. Obviously, there are chiral current loops circulating in the moiré cell, which generates large orbital magnetic moment in the TBG, resulting in large response of the valley splitting to the magnetic fields. It is worthwhile to note that the $C_{3z}$ symmetry is broken in the chiral current-loop pattern due to the presence of heterostrain in the TBG layers.

In Fig. 4b, we show the evolution of the calculated DOS of the TBG under vertical magnetic fields. Coulomb interactions have been included, and are treated within Hartree-Fock approximations at -1/2 filling as explained above. As the valence band of each valley carries nonzero Chern number, the occupied states below the chemical potential are actually orbital ferromagnetic states with nonvanishing net orbital magnetizations pointing along the $z$ direction. Supposing the magnetic field is along $z$, and the orbital magnetization is also aligned along $z$ by the external magnetic fields, then the correlated gap would be enhanced by increasing the magnetic fields. This is clearly shown in Fig. 4b and 4c. As the magnetic field increases from 0 to 5 T, the gap at -1/2 filling increases from 2 meV to 7 meV, which means there is a giant $g$ factor associated with orbital Zeeman effect. A straightforward linear fitting indicates that the effective $g$ factor is around 17.3, or the effective orbital magnetic moment in each moiré is about 8.7 $\mu_B$. This is qualitatively in agreement with our experimental results. Here, we should point out the splitting between the two valleys is much larger than the gap around the Fermi level (Fig. 4b). However, the slopes deduced from their variations with increasing the magnetic fields are the same.

In summary, we demonstrated that the *e-e* interactions are also very important even in the 1.68° TBG and a fully valley-polarized state is obtained when its VHS is half filled. Because the $C_{2z}$ symmetry breaking of the TBG induced by the substrate, our result indicated that there is large orbital magnetic moment in each moiré ~10.7 $\mu_B$ in the 1.68° TBG. Such a result demonstrated that the slightly TBG provides an unprecedented platform to explore magnetism that is purely orbital.

**References:**
1. Bistritzer, R. & MacDonald, A. H. Moire bands in twisted double-layer graphene. *Proc Natl Acad Sci* (*USA*) **108**, 12233-12237 (2011).
2. Xie, M., MacDonald, A. H. On the nature of correlated insulator states in twisted bilayer graphene. arXiv:1812.04213.

## Acknowledgements

This work was supported by the National Natural Science Foundation of China (Grant Nos. 11974050, 11674029). L.H. also acknowledges support from the National Program for Support of Top-notch Young Professionals, support from "the Fundamental Research Funds for the Central Universities", and support from "Chang Jiang Scholars Program".

## Author contributions

S.L. synthesized the samples, performed the STM experiments, and analyzed the data. J.L. performed the theoretical calculations. L.H. conceived and provided advice on the experiment, analysis, and the theoretical calculation. L.H., S.L., and J.L. wrote the paper. All authors participated in the data discussion.

**Competing financial interests**

The authors declare no competing financial interests.

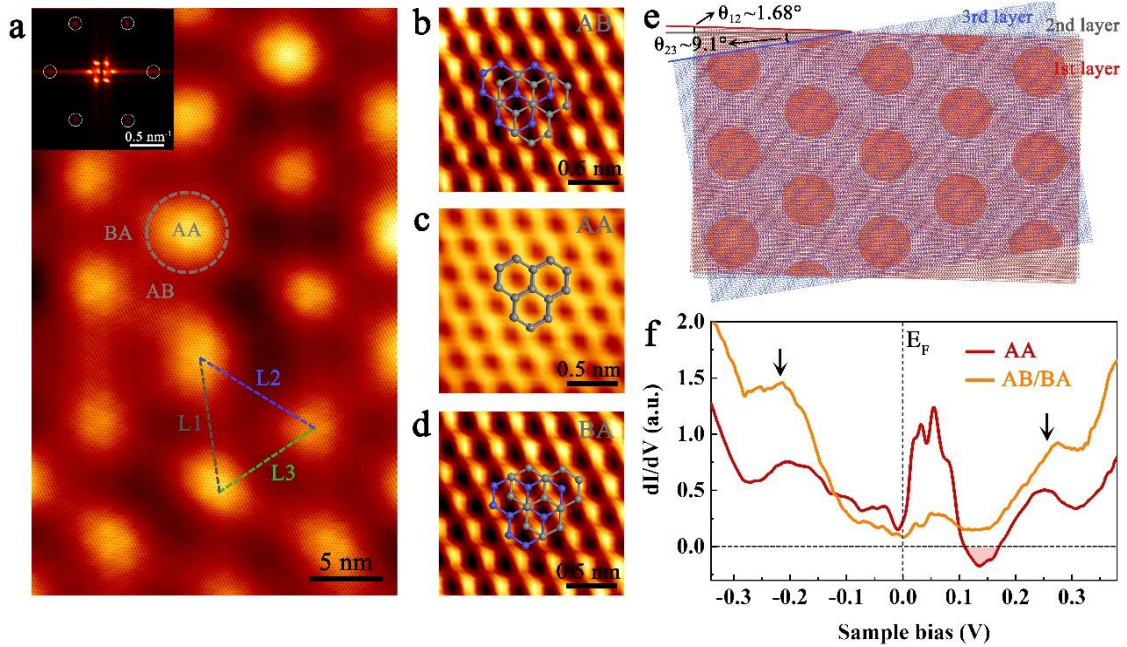

**FIG. 1. a.** A 25 nm ×40 nm STM image ($V_{sample}$ = 500 mV and $I$ = 0.3 nA) of multilayer graphene on Rh, which exhibits two series of moiré patterns with different periods. The inset shows FFT image of the moiré patterns. **b-d,** 2 nm ×2 nm atomic resolution STM images taken from AB-stacked region, AA-stacked region and BA-stacked region of the large-period moiré pattern. **e.** Schematic structure of the topmost twisted trilayer graphene. The twist angle between the first and the second layer is about 1.68 °, and the twist angle between the second and the third layer is about 9.1 °. **f.** dI/dV spectra taken from the AA-stacked region and AB/BA stacked region in the large-period moiré pattern. The position of Fermi level is marked. The low-energy VHSs in the AA-stacked region are completely empty. There is negative differential conductance between the low-energy VHSs and the remote bands.

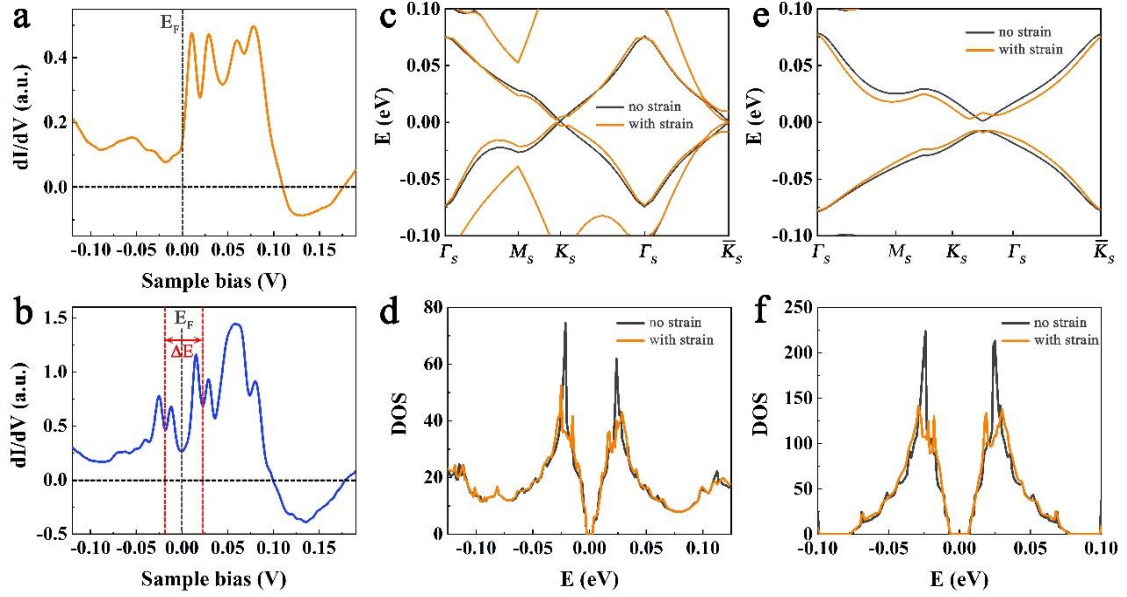

**FIG. 2. a and b.** High-resolution dI/dV spectra taken from the AA-stacked region of the 1.68 °TBG with different doping. In panel a, the low-energy VHSs are completely empty. In panel b, the left two VHSs are half filled and further split into four peaks. **c.** The bandstructures of the K valley in the 1.68 ° TBG supported by Bernal-stacked bilayer graphene calculated using the continuum Hamiltonian. The twist angle between the second layer and the third layer is 8.4 °. **d**. The corresponding DOS of the 1.68 °TBG supported by Bernal-stacked bilayer graphene. **e,f**. The bandstructures and corresponding DOS of the 1.68 °TBG with including the staggered sublattice potential. The blue and red lines indicate the energy bands without and with the uniaxial heterostrains.

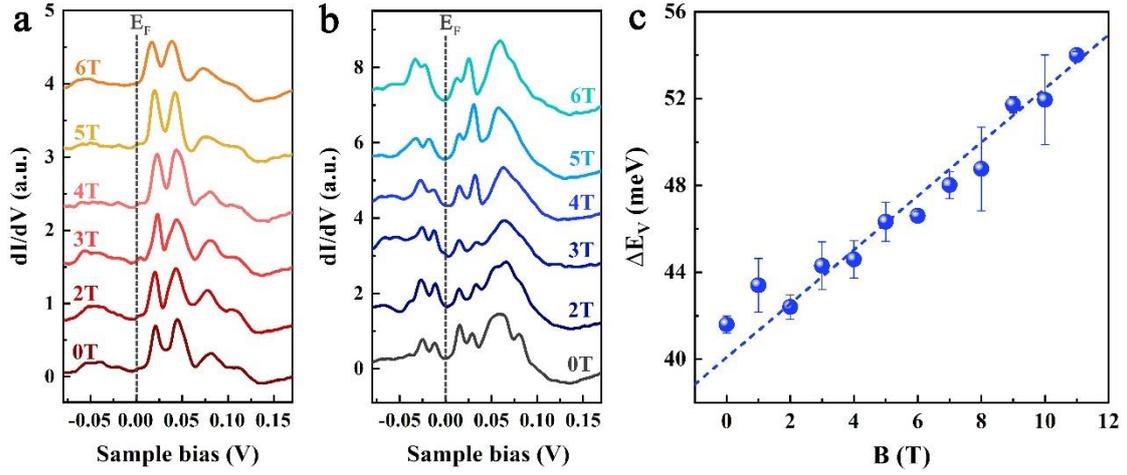

**FIG 3. a,b.** dI/dV spectra taken from the AA-stacked regions of the 1.68 °TBG with different doping as a function of magnetic fields. The positions of Fermi level are marked. **c.** Summarizing the splitting of the half-filled VHS in panel b as a function of magnetic fields. The blue dashed line marks the linear fitting between the valley splitting and the magnetic fields.

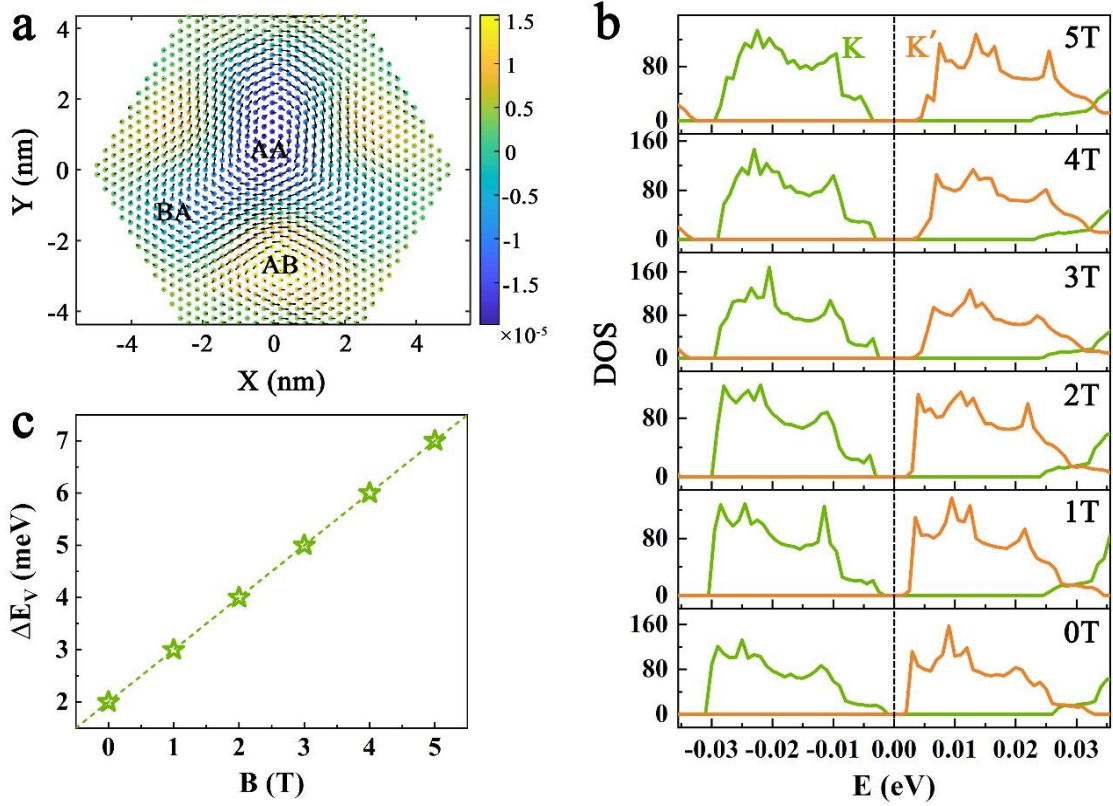

**FIG. 4. a.** The real-space distribution of the current density (black arrows) and the current-induced magnetic field (color coding) with the moiré supercell of the 1.68 °TBG. The AA region of the moiré pattern is centered at the origin. **b.** The evolution of the DOS of the 1.68 °TBG with including the staggered sublattice potential under different vertical magnetic fields. Coulomb interactions have been included, and are treated within Hartree-Fock approximations at -1/2 filling of the VHS. **c.** Linear fitting of the correlated gaps *vs.* the magnetic fields. The slope determines the effective *g* factor from the orbital magnetic moment.